\documentclass[pre,preprintnumbers,amsmath,amssymb]{revtex4}
\usepackage{graphicx}% Include figure files
\usepackage{bm}% bold math

\begin{document}
\title{Universal First-passage Properties of Discrete-time Random Walks
and L\'evy Flights on a Line: Statistics of the Global Maximum and Records}
\author {Satya N. Majumdar }

\affiliation{ Laboratoire de Physique Th\'eorique et Mod\`eles Statistiques,
        Universit\'e Paris-Sud. B\^at. 100. 91405 Orsay Cedex. France}  

\begin{abstract}
In these lecture notes I will discuss the universal
first-passage properties of a
simple correlated discrete-time sequence
$\{x_0=0, x_1,x_2,\ldots, x_n\}$ up to $n$ steps where $x_i$ represents
the position at step $i$ of a random walker hopping on a continuous line
by drawing independently, at each time step, a random jump length
from an arbitrary symmetric and continuous distribution (it includes, e.g.,
the L\'evy flights).
I will focus on the statistics of two extreme observables associated with the
sequence: (i) its global maximum and the time step at which the maximum occurs
and (ii) the number of records in the sequence and their ages. I will 
demonstrate
how the universal statistics of these observables emerge as 
a consequence of Pollaczek-Spitzer formula and the associated
Sparre Andersen theorem.

\end{abstract}

\maketitle

\date{\today}
\section{introduction}
Since the remarkable founding paper by Einstein in 1905~\cite{E1},
followed closely by two seminal papers respectively by
Smoluchowski~\cite{Smolu} and Langevin~\cite{Langevin},
random walks and the associated continuous-time Brownian motion
have remained
as fundamental cornerstones of statistical physics
with an amazingly impressive 
number of applications~\cite{Chandra,Feller,Hughes,MS,BKZ,BG} that 
range 
from traditional
`natural' sciences such as physics, chemistry, biology, mathematics
and astronomy  
all the way to `man-made' subjects such as computer science and finance.
Even though many aspects of this classical subject are extremely well understood
and form text book materials, it is fascinating that 
new questions with non trivial answers, arising from new applications,
continue to spring unexpected surprises.

In these lectures I will discuss some of these recent applications.
The general area of random walks is vast with an enormous literature. 
My goal for these lectures is rather modest. I will just focus
on a rather simple and restricted model: a discrete-time random hopper on
a continuous line. Starting from the origin $x_0=0$, the position of
the particle at step $n$ evolves via the Markov rule, 
$x_n=x_{n-1}+\xi_n$,
where $\xi_n$'s denote the random jumps at different time steps.
These jumps are independent and identically distributed (i.i.d.)
random variables, each drawn from the same distribution $\phi(\xi)$
which is symmetric and continuous. If the walk evolves up to step $n$,
one generates a sequence or a discrete-time series: $\{x_0=0, 
x_1,x_2,\ldots, x_n\}$. Clearly the members of this 
sequence are {\em correlated} random variables. 
Such a sequence is perhaps the simplest possible correlated
sequence that appears rather naturally    
in   
many different contexts. 
A classic
example of such a walk can be found in
bacterial   
chemotaxis, where a bacteria, in search of food,
jumps from one position to another at discrete time
steps~\cite{Koshland}.
In the context of
queueing theory $x_n$ represents the length of a single queue at time
$n$~\cite{queue}.
In the context of the evolution of stock prices $x_n$ represents the logarithm
of the price of a stock at time $n$~\cite{finance}. It can also
represent the $x$ coordinates of the beads of a Rouse polymer chain
in thermal equilibrium in $d$-dimensions (when the jump distribution is 
Gaussian)~\cite{Rouse} (see also \cite{CM}). When the jump distribution
has a power law tail with a divergent second moment, $\phi(\xi)\sim 
|\xi|^{-1-\mu}$
(with $0\le \mu\le 2$) for large $|\xi|$, 
this sequence
represents a L\'evy flight which also has enormous number
of applications~\cite{Levy1,Kinchin,Bertoin,BG,Metzler-Klafter1,Borovkov}.

Here I will focus on two {\em extreme} observables associated with such
a correlated sequence: (i) the global maximum $M_n={\rm 
max}\{0,x_1,x_2,\ldots,x_n\}$
of the sequence and the associated time step $m$ at which this maximum is realized in
a given sample (ii) the number and ages of records of this sequence where
a record is set to happen at step $i$ if $x_i$ is bigger than all
the previous values: $x_i> x_{k}$ for all $0\le k<i$. Age of a record
is simply the number of steps up to which this record survives, i.e, till
it gets surpassed by the next record breaking event.

Now, if the number of steps $n$ of the sequence is large and 
if the second moment of the jump length distribution $\sigma^2=\int_{\infty}^{\infty} 
\xi^2 \phi(\xi)\, d\xi$ is finite, one would expect, correctly, to 
recover the continuous-time limit results of the Brownian motion as a consequence
of the central limit theorem, at least for the global maximum ( 
records are not very well defined in the continuous-time limit). 
However, it turns out, as I will show here
in some detail, that many properties associated with extreme events
such as the global maximum or the number of records are {\em completely
universal for all} $n$, i.e., they do not depend on
the jump length distribution $\phi(\xi)$ at all whatever be the value of $n$, as long 
as
$\phi(\xi)$ is symmetric and continuous. Note, in particular, that this universality
does not even require a finite $\sigma^2$, e.g., it holds even for long range
L\'evy flights.

In fact, this universality has nothing to do
with the central limit theorem. Instead, it
will turn out to be a consequence of 
the Sparre Andersen theorem~\cite{SA} concerning the first-passage
properties of such a random walk sequence~\cite{Feller,Redner}. This is a rather
deep combinatorial theorem and the final result looks
deceptively simple though its derivation is far from simple.
Here I will provide a derivation of this result using
another result on the generating function
of the maximum of such a sequence, known as
the Pollaczek-Spitzer formula~\cite{Pollaczek,Spitzer}.
Somehow these results are not so well known among physicists.
So, I'll discuss these results in some detail and use them
to derive some universal and some nonuniversal properties
associated with the statistics of the maximum and the records
of this random walk sequence. 
  
In the latter half of my lectures in the school, I also discussed
the statistical properties of the functionals of Brownian motion
via the Feynman-Kac formula and in particular, various interesting
applications of the so called first-passage Brownian functionals, where
one considers the Brownian motion till its first-passage time.
They turn out to have various applications: in queueing theory,
in finance, in simple models of particle moving in a disordered
random potential and in astrophysics where one is interested
in the distribution of the life time of a comet in the solar system.
However, I will not include this interesting topic in these lecture notes, as I
have already discussed it in another article~\cite{review}. The interested
readers may consult this article and also another review
on Brownian functionals with interesting applications
in the localisation theory~\cite{CDT}.

This article is organised as follows. In Section II, I define
the model precisely and review some basic preliminaries to
remind the readers the central limit theorem and the L\'evy stable laws.
In Section III, I discuss the first-passage properties associated
with the random walk sequence and discuss the Pollaczek-Spitzer
formula and how this formula leads to the Sparre Andersen theorem.
Section IV is devoted to the statistics of the global maximum
and the universal statistics of the time of its occurrence where we use the Sparre 
Andersen theorem.
In Section V, we discuss the statistics of the number of records
and their ages and show how universal properties emerge again
as a consequence of the Sparre Andersen theorem. Finally, I 
conclude in Section VI with a summary and some open problems.

\section{Random Walks, Brownian Motion, L\'evy Flights: Some Preliminaries}

\subsection{Definitions}
Let us start with a simple discrete-time random walker moving on a continuous 
line. 
The
position $x_n$ of the walker after $n$ steps evolves for $n\ge 1$ via,
\begin{equation}
x_n = x_{n-1}+ \xi_n
\label{evol1}
\end{equation}
starting at $x_0=0$, where the step lengths $\xi_n$'s
are i.i.d. random variables
with zero mean and
each drawn from a normalized (to unity) distribution $\phi(\xi)$
which is symmetric,
$\phi(\xi)=\phi(-\xi)$ (see Fig. 1).
\vskip 0.2cm

Few examples of the jump length distribution $\phi(\xi)$ are:
\vskip 0.3cm

\noindent (i) $\phi(\xi)= \frac{1}{2}\, e^{-|\xi|}$ (Exponential) 

\vskip 0.3cm

\noindent (ii) $\phi(\xi)=\frac{1}{\sigma_0 \sqrt{2\pi}}\, 
e^{-\xi^2/{2\sigma_0^2}}$
 (Gaussian)

\vskip 0.3cm

\noindent (iii) $\phi(\xi)= \frac{1}{2}\left[\theta(\xi+1)-\theta(\xi-1)\right]$
(Uniform)

\vskip 0.3cm

\noindent (iv) $\phi(\xi) \sim |\xi|^{-1-\mu}$ for large $|\xi|$ with $0\le \mu\le 
2$
such that $\sigma^2=\int \xi^2\phi(\xi) d\xi$ does not exist (L\'evy flights)

\vskip 0.3cm

\noindent (v) $\phi(\xi)= \frac{1}{2}\delta(\xi+1) + \frac{1}{2}\delta(\xi-1)$ 
(Lattice random walk where the lattice spacing is unity).

\vskip 0.3cm

\noindent In first $4$ of these examples, the cumulative jump distribution
$\Psi(x)=\int_{-\infty}^{x} \phi(\xi)\, d\xi$ is a continuous function. In the last example (v), where
the walker is restricted to move on a one dimensional lattice
with unit lattice spacing, the cumulative jump 
distribution
$\Psi(\xi)$ is a non-continuous function. We will see later that
this continuity property of $\Psi(\xi)$ will play an important role.
Note further that in examples (i)-(iii) and (v), the variance
of the step length, $\sigma^2= \int_{-\infty}^{\infty} \xi^2\phi(\xi) d\xi$
is finite. We will see that in such cases the central limit theorem holds.
In the L\'evy case (iv), the central limit theorem breaks down.
\begin{figure}
\includegraphics[height=10.0cm,width=14.0cm,angle=0]{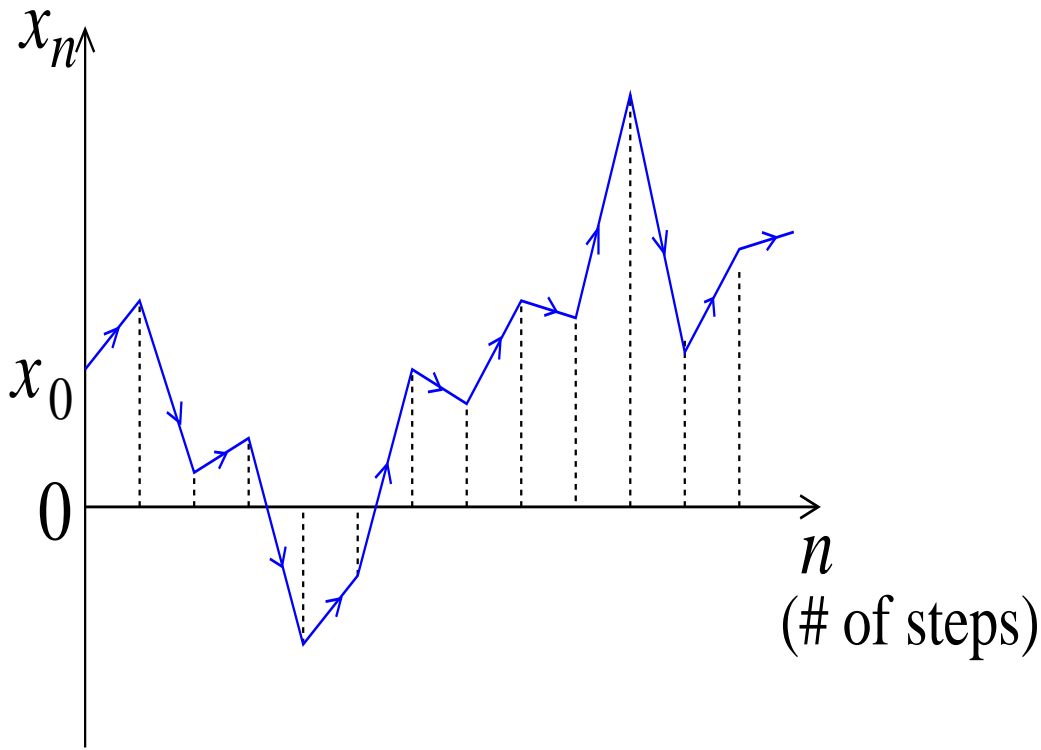}
\caption{\label{fig:rw1} A trajectory of a random walker
starting at the initial position $x_0$ and evolving with
the number of steps $n$.}
\end{figure}

\vskip 0.3cm 
The evolution equation \eqref{evol1} is Markovian since the
position $x_n$ at step $n$ depends only on the position at just the 
previous time step $x_{n-1}$ (and not on the full history before
the $(n-1)$-th step) and on the current 
noise, i.e., the noise $\xi_n$ at step $n$. This Markovian property
makes life simple as we will see later. As a simple example of
a non-Markovian evolution consider the rule
\begin{equation}
x_n= 2x_{n-1}-x_{n-2} + \xi_n
\label{nonmarkov}
\end{equation}  
where $\xi_n$'s are again i.i.d random variables. 
This is just the discrete-time version of the
continuous-time random acceleration problem: $d^2x/dt^2=\xi(t)$
where $\xi(t)$ is a Gaussian white noise with zero mean 
$\langle\xi(t)\rangle=0$ and delta correlator $\langle 
\xi(t)\xi(t')\rangle=\delta(t-t')$.
It turns out
that the first-passage properties of even this simple non-Markovian
evolution is highly nontrivial~\cite{sinai,burkhardt,persistence_review}. 
We will not consider non-Markov evolution rules further in these
lectures and focus only on the Markov evolution \eqref{evol1}. 
For the first-passage properties of non-Markovian
stochastic processes see ~\cite{persistence_review} and references
therein.

Iterating the Markov evolution rule \eqref{evol1} up to $n$ steps, it follows
that the position $x_n$ of the walker after $n$ steps, starting at $x_0=0$,
is simply a sum of $n$ i.i.d. random variables
\begin{equation}
x_n = \sum_{k=1}^n \xi_k.
\label{sum1}
\end{equation}
In the case when $\sigma^2$ is finite, using the independence property
of the step lengths $\xi_k$'s, it follows that the mean square displacement
of the particle after $n$ steps, for all $n$, is simply
\begin{equation}
\langle x_n^2\rangle = n\, \sigma^2.
\label{msd1}
\end{equation}

\vskip 0.3cm 
\noindent {\bf Brownian limit:}
At this point, it is useful to consider the continuous-time limit where
the random walk reduces to a Brownian motion. Let us define $\Delta t$
as a small time interval and set $t= n \Delta t$. Then \eqref{msd1} gives
\begin{equation}
\langle x_n^2\rangle = \frac{\sigma^2}{\Delta t}\, t.
\label{msd2}
\end{equation}
If one now takes the limit $\Delta t\to 0$, it follows that
$\sigma^2\to 0$ also in order that $\langle x^2(t)\rangle$ remains
finite at finite time $t$. Thus, to have a meaningful
continuous-time limit, the mean square step length $\sigma^2\to 2D \Delta t$
as $\Delta t \to 0$ with a finite diffusion constant $D$, leading to
the diffusive law of Brownian motion $\langle x^2(t)\rangle=2Dt$
for all $t$. In this continuous-time limit, one can also
rewrite the Markov evolution rule \eqref{evol1} as 
\begin{equation}
\frac{\Delta x_n}{\Delta t}= \frac{\xi_n}{\Delta t}= \xi(t)
\label{lange0}
\end{equation}
where $\xi(t)$ is random noise with zero mean that is uncorrelated
at two different times, $\langle \xi(t_1)\xi(t_2)\rangle = 0$ for $t_1\ne t_2$.
At the same time instant, however, $\langle \xi^2(t)\rangle = \sigma^2/(\Delta 
t)^2={2D}/{\Delta t}$. Thus, as $\Delta t\to 0$, $\langle \xi^2(t)\rangle$
diverges. A useful physicist's way of writing this correlation
function of the noise is $\langle \xi(t_1)\xi(t_2)\rangle = 2D 
\delta(t_1-t_2)$.
In this limit it is called the white noise and one writes \eqref{lange0}
as a stochastic Langevin equation
\begin{equation}
\frac{dx}{dt}= \xi(t)
\label{lange1}
\end{equation}
where $\xi(t)$ is the white noise with zero mean and a correlator
$\langle \xi(t_1)\xi(t_2) \rangle = 2D \delta(t_1-t_2)$.
Note that for all practical
purposes, such as in numerical simulation, one will interpret the delta 
function as $\delta (0)\equiv 1/{\Delta t}$.

We will see later that in the Brownian limit many properties of the walk,
such as its first-passage probability, become much simpler. In contrast, for
discrete time evolution, even though the process is Markov, some
of these properties are quite nontrivial.

\subsection{Green's Function}

Let us get back to our basic discrete-time Markov evolution \eqref{evol1}. 
In this subsection, let us compute
a basic object namely the free (bare) Green's function $G(x,x_0,n)$ defined
as the probability density of the position of the walker after step $n$
at $x$, given that 
it started from $x_0$ at step $0$. Using the Markov property, one can easily
write down a recursion relation for the evolution of $G(x,x_0,n)$
\begin{equation}
G(x,x_0,n)= \int_{-\infty}^{\infty} G(x',x_0, n-1)\,\phi(x-x')\, dx'
\label{forward1}
\end{equation}
which counts the event of particle jumping from its position $x'$ at step
$n-1$ to its position $x$ at step $n$ by an amount $(x-x')$ drawn from
the distribution $\phi(x-x')$. This is called the {\it forward} Kolmogorov
equation, since one considers the current position $x$ of the walker
as a variable. Alternatively, one can also write down a {\it backward}
Kolmogorov equation where one considers the starting position of the walker
$x_0$ as a variable
\begin{equation}
G(x,x_0,n)= \int_{-\infty}^{\infty} G(x, x_0', n-1)\, \phi(x_{0}'-x_0)\, 
dx_0'.
\label{backward1}
\end{equation}
Here one considers the displacement of the particle at the first step from
$x_0$ to $x_0'$ and for the subsequent evolution up to $(n-1)$ steps the
starting position of the walker is at $x_0'$. 
Both equations are completely equivalent to each other. We will see
later, however, that for certain first-passage related quantities, the
backward equation is often computationally more advantageous than the 
forward one.

These integral equations \eqref{forward1} or \eqref{backward1} can be
easily solved using Fourier transforms. For example, for the forward 
equation, we define 
\begin{equation}
{\tilde G} (k, x_0,n)= \int_{-\infty}^{\infty} G(x,x_0,n)\, e^{ikx}\, dx
\label{ft1}
\end{equation}  
and use the convolution form of \eqref{forward1} to get
${\tilde G}(k,x_0,n)= {\tilde G}(k,x_0,n-1)\,{\tilde \phi}(k)$
where ${\tilde \phi}(k)$ is the Fourier transform of $\phi(x)$.
Iterating $n$ times and using the initial condition, $G(x,x_0,0)=\delta(x-x_0)$
and hence ${\tilde G}(k,x_0,0)= e^{ikx_0}$ one gets
${\tilde G}(k,x_0,n)= \left[{\tilde \phi}(k)\right]^n e^{ikx_0}$.
Inverting the Fourier transform one obtains the exact Green's function
\begin{equation}
G(x,x_0,n)= \int_{-\infty}^{\infty} \left[{\tilde \phi}(k)\right]^n\, 
e^{-ik(x-x_0)}\, \frac{dk}{2\pi}
\label{green1}
\end{equation}
Let us now see what happens for large $n$. In cases where $\sigma^2$ is
finite, one has, for small $k$, ${\tilde \phi}(k) \approx 1- \sigma^2k^2/2 
+O(k^4)$. Now, for large $n$, the dominant contribution to the integral 
in \eqref{green1} comes from small $k$ region. Substituting the small
$k$ behavior, exponentiating and performing the Gaussian integral, one gets,
for large $n$, the standard Gaussian behavior
\begin{equation}
G(x,x_0,n) \simeq \frac{1}{\sqrt{2\pi n 
\sigma^2}}\,\exp\left[-\frac{(x-x_0)^2}{2n\sigma^2}\right]
\label{clt1}
\end{equation}
which is essentially the statement of the central limit theorem (CLT). 
Note that the universal Gaussian form holds only near the central peak but not
at the tails which are described by nonuniversal large deviation
function that I will not discuss here~\cite{Feller}.
On the other hand, for jump distributions with a divergent $\sigma^2$  
(such as for L\'evy flights in example (iv)), the CLT
breaks down~\cite{Levy1,Kinchin,BG}. For L\'evy flights, one can write for small 
$k$, ${\tilde 
\phi}(k)\approx 1- |a\,k|^{\mu}$ where 
$0\le \mu\le 2$ is the L\'evy index and $a$ is a microscopic length. 
Substituting this in \eqref{green1} and 
rescaling
$n (ak)^{\mu}\to k^{\mu}$ one gets, for large $n$,
\begin{equation}
G(x,x_0,n) \simeq \frac{1}{a\,n^{1/\mu}}\, 
\Phi_\mu\left(\frac{(x-x_0)}{a\,n^{1/\mu}}\right)
\label{levy1}
\end{equation}
where the function 
\begin{equation}
\Phi_\mu(z)=\int_{-\infty}^{\infty} e^{-|k|^{\mu}-ik z} \frac{dk}{2\pi}
\label{levy2}
\end{equation}
is called the L\'evy stable function of index 
$\mu$~\cite{Levy1,Kinchin,BG,Bertoin}.
Note that this function $\Phi_\mu(z)$, for large $z$, has the
same power law tail, $\Phi_{\mu}(z)\sim |z|^{-1-\mu}$ as
the jump distribution itself.
For some 
special
values of $\mu$, one can compute this function explicitly~\cite{BG}. Thus,
the result in \eqref{levy1} is the statement of L\'evy stable 
law~\cite{Levy1,Kinchin}:
the sum of i.i.d L\'evy distributed variables
is itself L\'evy distributed (up to a rescaling by $n^{1/\mu}$), i.e.,
the L\'evy distribution is stable under addition~\cite{Levy1,Kinchin,Bertoin}. 
This is thus the
counterpart of the CLT which is the analogous statement for
the sum of i.i.d random 
variables with a finite $\sigma^2$: the stable law for CLT is Gaussian. 
Note that from \eqref{levy1} it follows that the typical distance
traversed by the particle in step $n$ scales super-diffusively:
$x\sim n^{1/\mu}$ for $0<\mu\le 2$.

\vskip 0.3cm

\noindent {\bf Brownian limit:} In the continuous-time limit, when
$\sigma^2$ is finite and hence the CLT holds, the 
integral equations \eqref{forward1} or \eqref{backward1} reduce
to partial differential equations. For example, given the Langevin evolution 
in \eqref{lange1}, the forward Kolmogorov equation \eqref{forward1} reduces to
\begin{equation}
G(x,x_0, t+\Delta t) = \int_{-\infty}^{\infty} G(x-\xi(t) \Delta t, x_0, t)\, 
\phi(\xi(t))\, d\xi(t)
\label{brf1}
\end{equation} 
Expanding the Green's function on the rhs in a Taylor series, keeping terms up 
to $O((\Delta t)^2)$ and using the property that $\langle 
\xi(t)^2\rangle=\int_{-\infty}^{\infty} \xi^2\, \phi(\xi)\, d\xi= 2D/{\Delta 
t}$,
one gets, taking $\Delta t\to 0$, the well known diffusion equation for the 
Green's function
\begin{equation}
\frac{\partial G}{\partial t}= D \frac{\partial^2 G}{\partial x^2}
\label{diff1}
\end{equation}
starting from the initial condition, $G(x,x_0,0)=\delta(x-x_0)$.
Similarly one can write down a backward diffusion equation with
$x$ in \eqref{diff1} replaced by $x_0$. The solution
of the forward (or the backward)
diffusion equation can be easily found using Fourier transforms and one 
recovers,
as expected, the Gaussian behavior
\begin{equation}
G(x,x_0,t)= \frac{1}{\sqrt{4\pi D t}}\, 
\exp\left[-\frac{(x-x_0)^2}{4Dt}\right].
\label{gauss1}
\end{equation}

For L\'evy flights, where $\sigma^2$ is infinite and the CLT
breaks down, one can still {\em 
formally} define a continuous-time
limit, and obtain the so called L\'evy fractional diffusion 
equation (for a review and discussion, see ~\cite{Metzler-Klafter1}).
This simply follows by rewriting the basic recursion relation \eqref{forward1}
as
\begin{equation}
G(x,x_0,n)= \int_{-\infty}^{\infty} G(x-\xi_n,x_0,n-1)\,\phi(\xi_n)\, d\xi_n.
\label{levyf1}
\end{equation}
Next we write $G(x-\xi_n,x_0,n-1)= \int_{-\infty}^{\infty} {\tilde 
G}(k,x_0,n-1)\,e^{ik(x-\xi_n)}\,dk$
and substitute it in \eqref{levyf1}. This gives
\begin{equation}
G(x,x_0,n)= \int_{-\infty}^{\infty} dk\, {\tilde G}(k,x_0,n-1)\,{\tilde 
\phi}(k).
\label{levyf2}
\end{equation}
Following similar arguments as in the Brownian case, in the large $n$ limit,
one needs to keep only the small $k$ contribution of ${\tilde 
\phi}(k)=1-|ak|^{\mu}$
in \eqref{levyf2}. This gives
\begin{equation}
G(x,x_0,n)-G(x,x_0,n-1)\simeq - a^{\mu} \int_{-\infty}^{\infty} dk\,
|k|^{\mu}\,{\tilde 
G}(k,x_0,n-1).
\label{levyf3}
\end{equation}
Now, we need to divide both sides by the time increment $\Delta t$ and take
the limit $\Delta t \to 0$. To obtain a sensible limit, one needs to 
take $a\to 0$ limit as well, keeping the ratio $a^{\mu}/{\Delta t}=K$ fixed.
This gives a continuous-time integro-differential equation
\begin{equation}
\frac{\partial G}{\partial t}= -K \int_{-\infty}^{\infty} dk\, {\tilde
G}(k,x_0,n-1)\, |k|^{\mu}= -K \left(-\partial_x^2\right)^{\mu/2}G
\label{levyf4}
\end{equation}
where the integral in the $k$ space can be {\em formally}
interpreted as a fractional derivative. Note that for $\mu=2$,
one recovers the standard diffusion equation. But for $0\le \mu<2$,
one still needs to solve an integral equation even
in the continuous-time limit. Thus for the L\'evy walks, even though
one can formally write down a continuous-time equation, it is not
as useful as the ordinary Brownian case where one has a true differential
equation in real space whose solution can be easily obtained. 
This continuous-time fractional diffusion equation has
been studied extensively in the recent past (for a review
see \cite{Metzler-Klafter1}) and many interesting results,
in particular concerning first-passage properties
have been derived (see for instance
\cite{Koren,ZRK}). However, in these lectures I will not 
use this approach and
will rather stick to the discrete-time evolution.

\section{Random Walks: Survival and First-passage}

Having done with these standard basic preliminaries, let us now turn to the 
first-passage properties of a random walk evolving in discrete time
via the Markov rule \eqref{evol1} with arbitrary symmetric jump length 
distribution $\phi(\xi)$. We first define the restricted Green's function
$G^{+}(x,x_0,n)$ as the probability (density) that the walker, starting at 
$x_0>0$
at step $0$, reaches the position $x>0$ at step $n$ but {\it without crossing
the origin} in between, i.e., it stays positive at all intermediate steps
and lands at $x$ exactly at the $n$-th step
\begin{equation}
G^{+}(x,x_0,n)= {\rm Prob}\left[x_n=x,\, x_{n-1}\ge 0,\,x_{n-2}\ge 
0,\ldots,x_1\ge 0|x_0\right].
\label{rgreen}
\end{equation}
Using the Markov
property of the evolution, one can again write down the evolution
equation for the restricted Green's function, both {\it forward}
and {\it backward} as in case of free Green's function in the
previous subsection
\begin{eqnarray}
G^{+}(x,x_0,n) &=& \int_0^{\infty} G^{+}(x',x_0,n-1)\, \phi(x-x')\, dx' ;\quad\, 
{\rm (forward)} \label{rf1} \\
G^{+}(x,x_0,n) &=& \int_0^{\infty} G^{+}(x,x_0',n-1)\, \phi(x_0'-x_0)\, dx_0'; 
\quad
{\rm (backward)} \label{rb1}  
\end{eqnarray}
The interpretation is as before. For example, in the forward case, one
considers the walker reaching $x'$ at step $(n-1)$ (staying positive always)
and then making a final jump $x'\to x$ at step $n$ by drawing
a random length $x-x'$ from the distribution $\phi(\xi)$.
Similarly, in the backward equation, the particle at step $1$ jumps
from its initial position $x_0$ to a new position $x_0'$ and
subsequently evolves for $(n-1)$ steps starting from this new initial
position $x_0'$ while staying positive all along. One then integrates over all
possible jumps at the first step, but making sure that $x_0'$ is positive.

The {\it survival probability} or the {\it persistence} is defined
as the probability $Q(x_0,n)$ that the particle survives (i.e. stays 
positive) up to step $n$, no matter what the final position $x$ at step $n$ 
is. Thus
\begin{equation}
{\rm Survival \,\,\, Probability:}\quad \quad Q(x_0,n)= 
{\rm Prob}\left[x_n\ge 0,\,x_{n-1}\ge 0,\,x_{n-2}\ge 0,\ldots, x_1\ge 
0|x_0\right]=
\int_0^{\infty} 
G^{+}(x,x_0,n)\, dx
\label{pers1}
\end{equation}
Thus, one can either solve first the forward equation \eqref{rf1}, obtain
the restricted Green's function $G^{+}(x,x_0,n)$ for all $x$ and then
integrate over $x$ in \eqref{pers1} to obtain the survival probability
$Q(x_0,n)$. Alternatively, and in a much easier way, one can
integrate the backward equation \eqref{rb1} over $x$ and write directly
a backward evolution equation for the survival probability itself
\begin{equation}
Q(x_0,n)= \int_0^{\infty} Q(x_0', n-1)\, \phi(x_0'-x_0)\, dx_0'.
\label{pers2}
\end{equation}
Thus one saves an extra integration step \eqref{pers1} and just needs to solve 
only 
the integral equation \eqref{pers2} starting from the initial
condition $Q(x_0,0)=1$ for all $x_0\ge 0$. This initial
condition follows from the fact that the walker
definitely (with probability $1$) does not cross $0$ in $0$ step.
One thus sees why the backward equation \eqref{pers2} is more 
advantageous compared to the forward equation, atleast as far
as the persistence properties are concerned.

Once we have obtained the survival probability $Q(x_0,n)$, the
{\it first-passage probability} can be easily
computed from it. The first-passage probability
$F(x_0,n)$ is defined as the probability 
that the walker, starting initially at $x_0$,  {\it crosses the origin for the 
first time} immediately after
step $n-1$ (i.e., it is positive at step $n-1$, but becomes negative at step 
$n$). It then follows that 
\begin{equation}
F(x_0,n)= Q(x_0,n-1)- Q(x_0,n)
\label{fp1}
\end{equation}
as it counts the fraction of paths that survived up to step $(n-1)$, but
not up to step $n$. 

So, to compute the first-passage or the survival probability, we need
to solve the integral equations \eqref{rf1}, \eqref{rb1} or just
directly \eqref{pers2}.
Note the important differences in these equations
compared to the free Green's functions in \eqref{forward1}
and \eqref{backward1}: they look almost similar, but not quite. In
equations \eqref{rf1}, \eqref{rb1} or \eqref{pers2}, the limit of integration
on the rhs is from $0$ to $\infty$, as opposed to $-\infty$ to $\infty$
in the free Green's function equations \eqref{forward1} and \eqref{backward1}. 
This makes a huge difference! The reason is, even though 
\eqref{pers2} apparently seems to have a convolution form, the limit
of integration is only over half-space $[0,\infty]$ and not the full
space $[-\infty,\infty]$. If the limits were over the full space, as in
the case of free Green's functions, one can simply use the Fourier transform 
methods. But for the half-space problem, unfortunately one can not use
simple Fourier transform technique. In fact, such half-space integral
equations have been well studied in mathematics and are known as
Wiener-Hopf integral equations~\cite{MF}. For a general kernel $\phi(x-x')$,
they are notoriously difficult to solve! However, for the particular
case where the kernel $\phi(x-x')$ has the interpretation of a probability
density function (i.e., non-negative and normalizable  function), one can
obtain explicit solution~\cite{Spitzer} (as discussed later). 

The discussion above makes it clear the technical
reason as to why computing the 
first-passage 
properties of even a simple random walker (but with arbitrary jump 
distribution $\phi(\xi)$) is nontrivial. Before we write the solution
explicitly, let us see first how this problem simplifies in
the continuous-time Brownian limit.

\vskip 0.3cm

\noindent {\bf Brownian limit:} In the Brownian limit, one can
reduce the discrete time backward {\it integral} equation \eqref{pers2} for 
the
survival probability into a partial differential equation. Let us consider
the survival probability $Q(x_0, t+\Delta t)$ up to time $t+\Delta t$.
Let us break the interval $[0,t+\Delta t]$ into two intervals
$[0,\Delta t]$ and $[\Delta t, t+ \Delta t]$. In the first small
interval the particle evolves from its initial position $x_0$ to
a new random position $x_0+ \xi(0)\Delta t$ where $\xi(0)$ is the
initial noise in the Langevin equation \eqref{lange1}. Subsequently
the particle evolves in the interval $[\Delta t,  t+ \Delta t]$ starting
from its new initial position $x_0+ \xi(0)\Delta t$. Thus, the analogue
of \eqref{pers2} is
\begin{equation}
Q(x_0, t+\Delta t)= \int_0^{\infty} Q(x_0+ \xi(0) \Delta t, t)\,\phi(\xi(0))\, 
d(\xi(0))
\label{bfp1}
\end{equation}
Expanding in a Taylor series as in the case of free Green's function and using
the properties of the white noise, one then gets the backward Fokker-Planck 
equation for the survival probability
\begin{equation}
\frac{\partial Q}{\partial t}= D \frac{\partial^2 Q}{\partial x_0^2}
\label{bfp2}
\end{equation}
valid for all $x_0\ge 0$ and to be solved with the boundary conditions:
(a) $Q(x_0=0,t)=0$ for all $t$ and (b) $Q(x_0\to \infty, t)=1$ for all $t$
and subject to the initial condition $Q(x_0,0)=1$ for all $x_0>0$. Thus
in the Brownian limit, we are able to reduce the Wiener-Hopf integral
equation into a partial differential equation (PDE): that's already
a big simplification! 

The solution to this PDE can be obtained
by various standard methods. Let me just mention here a
slightly non-standard but quick method. Clearly, using
the diffusive scaling $x\sim t^{1/2}$, it follows that
the function $Q(x_0,t)$ must have a scaling form: 
$Q(x_0,t)=U\left(\frac{x_0}{\sqrt{4Dt}}\right)$.
Substituting this scaling form in the PDE \eqref{bfp2} one
obtains an ordinary differential equation (that's what scaling
always does: reduces a function of two variables into a 
function of a single scaled variable) valid for 
$z\ge 0$
\begin{equation}
\frac{d^2 U}{dz^2} + 2\, z\, \frac{dU}{dz}=0.
\label{ode1}
\end{equation}
The initial and the boundary conditions of the PDE translates
into two boundary conditions: $U(0)=0$ and $U(z\to \infty)=1$.
The solution can be easily obtained: $U(z) = {\rm erf}(z)= 
\frac{2}{\sqrt{\pi}}\,\int_0^{z} e^{-u^2}\, du$. 
Thus we get the explicit well known solution~\cite{Feller,Redner}
\begin{equation}
Q(x_0,t) = {\rm erf}\left(\frac{x_0}{\sqrt{4Dt}}\right)
\label{erf1}
\end{equation}
Note that even though we had assumed scaling (without really proving it!),
the solution \eqref{erf1} is exact for all $t$ as one can directly verify 
by substituting it in the PDE \eqref{bfp2}.
One also sees that for large $t$ and fixed $x_0$  the survival probability
decays as a power law
\begin{equation}
Q(x_0,t) \simeq \frac{x_0}{\sqrt{\pi D\, t}}.
\label{pers3}
\end{equation}
The first-passage probability is given by $F(x_0,t)= -\frac{\partial 
Q}{\partial t}$ which is just the continuous-time limit of \eqref{fp1}.
Using \eqref{erf1} one then gets
\begin{equation}
F(x_0,t) = \frac{x_0}{\sqrt{4\pi D t^3}}\,e^{-x_0^2/{4Dt}}
\label{fp2}
\end{equation}
which decays, for large $t$ and fixed $x_0$, as $t^{-3/2}$ with
the famous first-passage exponent $3/2$~\cite{Chandra,Feller,Redner}.

\subsection{Pollaczek-Spitzer formula and Sparre Andersen Theorem}

Let us now go back to the basic Wiener-Hopf integral equation \eqref{pers2}
that
describes the evolution of the survival probability $Q(x_0,n)$. As mentioned
before, the solution is nontrivial for a general kernel $\phi(x-x')$.
However, when the cumulative distribution $\Psi(x)=\int_{-\infty}^x 
\phi(\xi)d\xi$ is a continuous function
such as in examples (i)-(iv) in Section-I (but not for lattice random walk
(v) where $\Psi(x)$ is a discontinuous function), an explicit solution 
was first found by Pollaczek~\cite{Pollaczek} and later independently by 
Spitzer~\cite{Spitzer} in a slightly
different context. Pollaczek was interested in finding the distribution
of the ordered partial sums of a set of i.i.d. variables, whereas Spitzer
was interested in finding the distribution of the maximum of the
set of partial sums, which is related (see later) to the survival
probability. Spitzer's derivation was more combinatorial. The
same integral equation also appeared previously in a variety
of half-space transport problems in physics and astrophysics (see
\cite{Ivanov} and references therein) and several other derivations
of the solution of this equation,
mostly algebraic in nature, are known~\cite{Ivanov}.
Unfortunately, all these derivations, both the combinatorial as well
as the algebraic ones, are highly technical in nature and
there is no easy way! Here I will avoid these technical
steps and instead just state the final result and discuss
its applications. Readers who are interested in the algebraic
derivation
may consult ~\cite{MCZ} where we have listed systematically
the steps that lead to the final solution. 

The solution
of \eqref{pers2}, with the initial condition $Q(x_0,0)=1$ for all $x_0>0$, 
is in terms of a double Laplace transform of $Q(x_0,n)$
\begin{equation}
\int_0^{\infty} \left[\sum_{n=0}^{\infty} Q(x_0,n)\, s^n\right]\, 
e^{-p\,x_0}\, dx_0= \frac{1}{p\sqrt{1-s}}\, \exp\left[-\frac{p}{\pi}
\int_0^{\infty} \frac{\ln \left(1-s {\tilde \phi}(k)\right)}{p^2+k^2}\, dk 
\right] 
\label{psp1}
\end{equation}
where ${\tilde \phi}(k)= \int_{-\infty}^{\infty} \phi(\xi)\, e^{ik\xi}\, d\xi$
is the Fourier transform of the jump length distribution. We will refer
this solution in \eqref{psp1} as the Pollaczek-Spitzer formula.

Let us now discuss some consequences of this explicit result.

\subsection{Sparre Andersen Theorem}

Although the survival probability $Q(x_0,n)$ for arbitrary $x_0$ depends
explicitly on the jump length distribution $\phi(\xi)$ as evident in 
\eqref{psp1}, it turns out that $Q(0,n)$ (the survival probability
of the particle up to $n$ steps starting at the origin) becomes, somewhat
miraculously, {\it independent} of the distribution $\phi(\xi)$ as long as
it is a continuous function. To see this, let us take the $p\to \infty$ limit
in \eqref{psp1}. Making a change of variable $px_0=y$ on the lhs of 
\eqref{psp1} and taking $p\to 
\infty$ limit, the lhs reduces, to leading order, to $\frac{1}{p}\, 
\sum_{n=0}^{\infty} Q(0,n)\, s^n$. On the rhs, taking $p\to \infty$ limit
gives $\frac{1}{p\sqrt{1-s}}$. Equating the leading order terms 
(of $O(1/p)$ for large $p$) on both sides
gives the identity, for all $s$,
\begin{equation}
\sum_{n=0}^{\infty} Q(0,n)\, s^n = \frac{1}{\sqrt{1-s}}.
\label{SA1}
\end{equation}
Equating powers of $s$ one gets the Sparre Andersen theorem~\cite{SA}
\begin{equation}
q(n)= Q(0,n)= {{2\,n} \choose {n}}\, 2^{-2\,n}
\label{SA2}
\end{equation}
where we have used, for convenience, a shorthand notation $q(n)$ for $Q(0,n)$.
Thus, quite amazingly, the survival probability $q(n)=Q(0,n)$ (starting
from the origin) is completely {\it universal} and that too {\it for all}
$n$ (and not just for large $n$). No matter whether the jump length 
distribution is exponential, Gaussian or uniform, $q(n)$ is the same
and is given by the simple formula in \eqref{SA2}.
Sparre Andersen derived this formula originally using rather
involved combinatorial approach. 
This simple looking formula is however a bit deceptive and led several
authors to try to derive it in a `simple' way! Unfortunately, all
attempts led to equally complicated derivation (see \cite{FF} and
references therein). Deriving this
formula as a special case of the Pollaczek-Spitzer solution is
instructive as it shows that the role of the starting point $x_0=0$
is important for this {\it universality}. One looses this universality
the moment $x_0$ is nonzero.

Let us also note another interesting fact. In the limit of large $n$,
the survival probability $q(n)$ in \eqref{SA2} decays, to leading order, as
\begin{equation}
q(n)=Q(0,n) \simeq \frac{1}{\sqrt{\pi n}}.
\label{pspasymp}
\end{equation}
Let us emphasise again that this result holds for arbitrary continuous
jump distribution $\phi(\xi)$ including even the L\'evy flights! 
One may `naively' remark that this $n^{-1/2}$ asymptotic decay is equivalent to 
the $t^{-1/2}$ decay
of the survival probability in the Brownian limit derived in
\eqref{pers3}. However, this is not correct and is actually
rather subtle as was shown in~\cite{MCZ}. 
Consider first a continuous and symmetric jump distribution
with a finite second moment $\sigma^2=\int_{-\infty}^{\infty}\xi^2\, 
\phi(\xi)\, d\xi$. 
To derive the Brownian limit from the Pollaczek-Spitzer formula \eqref{psp1},
one first considers the scaling limit $x_0\to \infty$ and $n\to \infty$
but keeping the ratio $x_0/\sqrt{n}$ fixed. A careful asymptotic analysis
of \eqref{psp1} shows that in this limit the first two leading terms for large 
$n$ are given by~\cite{MCZ}
\begin{equation}
Q(x_0,n) \simeq {\rm erf}\left(\frac{x_0}{\sqrt{2\sigma^2 n}}\right)+ 
\frac{1}{\sqrt{\pi n}}\, e^{-x_0^2/{2\sigma^2 n}} 
\label{pspasymp1}
\end{equation}
If one now takes the $x_0<<\sqrt{n}$ limit, one recovers the
universal Sparre Andersen result in \eqref{pspasymp} from the 
second term on the rhs of Eq. (\ref{pspasymp1}). On the
other hand, if one keeps the scaling ratio $x_0/\sqrt{n}$ fixed
and takes the strict $n\to \infty$ limit, the second term
in \eqref{pspasymp1} becomes subleading and the first term
on the rhs (which remains nonuniversal in this limit as it contains
$\sigma^2$ explicitly) becomes the leading term that
provides the Brownian result in
\eqref{erf1} upon identifying $\sigma^2 n= 2Dt$.
Thus the $n^{-1/2}$ universal decay of the survival probability
(for $x_0=0$) is not quite related to the Brownian result $t^{-1/2}$:
they originate from two different terms in \eqref{pspasymp1}.

\vskip 0.3cm

\noindent {\bf Generalization to asymmetric jump distribution:} Actually there 
exists a 
generalized Sparre Andersen theorem~\cite{SA} which
holds for non-symmetric (but still continuous) jump length distribution 
$\phi(\xi)$. Unlike in the symmetric case, for asymmetric jump distribution
of a random walk starting at $x_0=0$,
the probability that the walker is on the positive side up to $n$ steps
is different from the probability that it is on the negative side
up to $n$ steps. Thus one needs to define two different survival probabilities 
\begin{eqnarray}
q_{+}(n) &=& {\rm Prob}[x_n\ge 0,\, x_{n-1}\ge 0,\ldots, x_1\ge 0|x_0=0] 
\label{upsurv} \\
q_{-}(n) &=& {\rm Prob}[x_n\le 0,\, x_{n-1}\le 0,\ldots, x_1\le 0|x_0=0]
\label{downsurv}
\end{eqnarray}
For symmetric jump distribution $q_{+}(n)=q_{-}(n)=q(n)$.
In the asymmetric case, the generalized Sparre Andersen theorem reads
\begin{eqnarray}
{\tilde q}_{+}(s)&=&\sum_{n=0}^{\infty} q_{+}(n)\, s^n = 
\exp\left[\sum_{n=1}^{\infty} 
\frac{p^{+}_n}{n}\,s^n\right] \label{gsaup} \\
{\tilde q}_{-}(s)&=& \sum_{n=0}^{\infty} q_{-}(n)\, s^n = 
\exp\left[\sum_{n=1}^{\infty}
\frac{p^{-}_n}{n}\,s^n\right]
\label{gsadown}
\end{eqnarray}
where $p^{+}_n= {\rm Prob}(x_n\ge 0)=\int_0^{\infty} G(x,x_0,n)\,dx $ 
and $p^{-}_{n}={\rm Prob}(x_n\le 0)=\int_{-\infty}^{0} G(x,x_0,n)$
are just the 
probabilities
that exactly at the
$n$-th step the particle position is positive and negative respectively. For the 
symmetric (zero bias)
case, $p^{+}_n=p^{-}_n=1/2$ (by symmetry) and then both
equations \eqref{gsaup} and \eqref{gsadown} reduce to 
\eqref{SA1}.

Let us mention here a special case with drift, noted by Le Doussal and 
Wiese~\cite{LW}, 
that is explicitly solvable and
that gives rise to a power law decay of the survival probability
with a continuously dependent exponent. Consider the evolution,
\begin{equation}
x_n= x_{n-1}+ \mu +\xi_n
\label{drift}
\end{equation}
with $x_0=0$.
Here $\mu$ represents a drift and $\xi_n$'s
are i.i.d noise variables each drawn from a symmetric Cauchy distribution
\begin{equation}
\phi(\xi)= \frac{a}{\pi (\xi^2+a^2)}
\label{cauchy1}
\end{equation}
In this case, the variable $y_n=x_n-\mu\, n$ undergoes a symmetric random 
walk,
$y_n=y_{n-1}+\xi_n$. Hence, the probability distribution of $y_n$
at step $n$, starting from $y_0=0$, can be easily computed from
the free Green's function discussed in Section I. In fact, the Cauchy 
distribution corresponds to the L\'evy laws in \eqref{levy1}
with index $\mu=1$. Hence, 
\begin{equation}
G(y, 0, n)= \frac{1}{a\,n}\, \Phi_1\left(\frac{y}{a\,n}\right)= \frac{an}{\pi 
(y^2+a^2 n^2)}
\label{cauchy2}
\end{equation}
Thus
\begin{eqnarray}
p^{+}_n &=& {\rm Prob}(x_n\ge 0)= {\rm Prob}(y_n\ge -\mu\, n)= \int_{-\mu\, 
n}^{\infty} 
\frac{an}{\pi
(y^2+a^2 n^2)}\, dy= \frac{1}{2}+\frac{1}{\pi}\tan^{-1}(\mu/a) 
\label{cauchy31} \\
p^{-}_n &=&  {\rm Prob}(x_n\le 0)= {\rm Prob}(y_n\le -\mu n)=
 \int_{-\infty}^{-\mu\, n}  
\frac{an}{\pi
(y^2+a^2 n^2)}\, dy = \frac{1}{2}-\frac{1}{\pi}\tan^{-1}(\mu/a)
\label{cauchy32}
\end{eqnarray}
Substituting these results in \eqref{gsaup} and \eqref{gsadown} one gets
\begin{equation}
{\tilde q}_{\pm}(s)=\sum_{n=0}^{\infty} q_{\pm}(n)\, s^n = 
\frac{1}{(1-s)^{\zeta_{\pm}}};\quad\, 
\zeta_{\pm}=\frac{1}{2}\pm \frac{1}{\pi}\tan^{-1}(\mu/a).
\label{cauchy4}
\end{equation}
Inverting the generating function one then finds that for large $n$
\begin{equation}
q_{\pm}(n) \simeq \frac{1}{\Gamma(\zeta_{\pm})}\, 
\frac{1}{n^{\theta_{\pm}(\mu)}};\quad\, 
\theta_{\pm}(\mu)=1-\zeta_{\pm}= \frac{1}{2}\mp \frac{1}{\pi}\tan^{-1}(\mu/a).
\label{cauchy5}
\end{equation}
Thus the persistence exponents $\theta_{\pm}(\mu)$ are nontrivial and vary 
continuously with 
the drift $\mu$. For example, as $\mu\to \infty$ (drift away from the origin), 
$\theta_{+}\to 
0$ (the particle always 
remains positive) and as $\mu\to -\infty$ (drift towards the origin), 
$\theta_{+}\to 1$
leading to a faster decay than the driftless ($\mu=0$) case where 
$\theta_{\pm}(0)=1/2$.

\section{First Application: Statistics of the Maximum of the Walk}

The study of the statistics of the maximum of a set of i.i.d. random variables
goes back a long way and the subject is called Extreme Value 
Statistics (EVS)~\cite{Gumbel}. The results are well established and have found a lot 
of 
applications in
a wide variety of fields~\cite{Gumbel}. However, the standard EVS, developed for 
i.i.d.
variables, does not apply when the random variables are {\em correlated}.
Recently there has been growing
interests in the statistics of the maximum of a set
of {\em correlated} random variables~\cite{Krapiv}. The random walk model
discussed in this article presents a solvable example
of the statistics of maximum of a set of strongly correlated variables.

More precisely, let us consider again the sequence \eqref{evol1}
starting from $x_0=0$ and the successive noise variables $\xi_k$'s are
as usual i.i.d variables each drawn from a symmetric and continuous
$\phi(\xi)$.
Let us define the global maximum of the walk up to $n$ steps
\begin{equation}
M_n= {\rm max}(0,x_1,x_2,\ldots, x_n).
\label{max1}
\end{equation}
Clearly $M_n$ is a random variable taking different values for different
realizations of the walk and we would like to compute the distribution
of $M_n$.
Note that even though the noise variables $\xi_k$'s are uncorrelated, the 
position
of the walker $x_k$'s are correlated. For example, when $\sigma^2=\langle 
\xi^2\rangle$ is finite, it is easy to see from \eqref{evol1} that
\begin{equation}
\langle x_m x_n\rangle = \sigma^2\, {\rm min}(m,n) 
\label{corr1}
\end{equation}
Thus, this is clearly an example where one is trying to compute
the distribution of a set of correlated random variables.

The distribution of $M_n$, as we will see now, is actually closely
related to the survival probability $Q(x_0,n)$ discussed in the
previous section. To establish this connection, let us first define
the cumulative distribution ${\rm Prob}(M_n\le y)$. This is just
the probability that the walk, starting at $x_0=0$ at step $0$, stays
below the level $x=y$ up to step $n$, i.e.,
\begin{equation}
{\rm Prob}(M_n\le y)= {\rm Prob}\left[x_1\le y,\, x_2\le y,\,\ldots, x_n\le 
y\right].
\label{max2}
\end{equation}
Let us make a shift and define $z_k=y-x_k$. 
Then,
$z_k$'s evolve via the same Markov rule \eqref{evol1}, but
starting from the initial position $z_0=y$ (since $x_0=0$).
Thus \eqref{max2} reduces to
\begin{equation}
{\rm Prob}(M_n\le y)={\rm Prob}\left[z_1\ge 0,\, z_2\ge 0,\,\ldots, z_n\ge 
0|z_0=y\right]= Q(y,n)
\label{max3}
\end{equation}
where $Q(y,n)$ is precisely the survival probability of the walk up to $n$ 
steps, starting at $y$. The solution of $Q(y,n)$ is given by
the Pollaczek-Spitzer formula \eqref{psp1} for arbitrary continuous 
distribution $\phi(\xi)$. 

\subsection{Expected Maximum}

The exact solution for $Q(y,n)$ in \eqref{psp1}
thus also provides an exact solution (or rather the double Laplace transform)
of the probability distribution of the maximum, at least in principle.
In practice however, the extraction of the moments of the maximum
from this explicit Pollaczek-Spitzer formula \eqref{psp1} turns out
to be rather nontrivial. 
For instance, even the first moment, i.e., the expected maximum $
E[M_n]$ 
is hard to extract for all $n$ and arbitrary continuous noise 
distribution $\phi(\xi)$. This question first arose in the context of 
a packing problem in two dimensions
where $n$ rectangles of variable sizes
are packed in a semi-infinite strip of width one~\cite{CS,CFFH}.
It was shown in Ref. \cite{CFFH} that for the special case of the 
uniform jump distribution,
$\phi(\xi)=1/2$ for $-1\le \xi \le 1$ and $\phi(\xi)=0$ outside, for large 
$n$,
\begin{equation}
E[M_n] = \sqrt{\frac{2n}{3\pi}} -0.297952\dots + O(n^{-1/2}).
\label{flaj1} 
\end{equation}
The leading $\sqrt{n}$ behavior is easy to understand and can be 
derived from the corresponding
behavior of a continuous-time Brownian motion after a 
suitable rescaling~\cite{CFFH}. However,
the leading finite-size correction term turns out to be a nontrivial constant 
$-c$ with
$c=0.29795219028\dots$ that
was computed in Ref.\ \cite{CFFH} by 
enumerating an intricate double series obtained after
a lengthy calculation by a different method.
It is important to compute the leading finite size correction
term very precisely as it provides a sharper estimate of
the efficiency of rectangle packing algorithms studied in computer
science~\cite{CS,CFFH}.  

Recently, we were able to show~\cite{CM}, starting from
the Pollaczek-Spitzer formula \eqref{psp1}, that for   
arbitrary continuous and symmetric
jump
distribution 
$\phi(\xi)$ with a finite second moment $\sigma^2=\int_{-\infty}^{\infty} 
\xi^2\, \phi(\xi)\, d\xi$,
the expected
maximum has a similar asymptotic behavior as in the uniform case, namely,
\begin{equation}
E[M_n]= \sigma \sqrt{\frac{2n}{\pi}} - c + O(n^{-1/2}).
\label{cm1}
\end{equation}
Moreover, an exact expression for the constant $c$ was found~\cite{CM}
\begin{equation}
c = - \frac{1}{\pi}\, \int_0^{\infty} \frac{dk}{k^2}\, \ln 
\left[\frac{1-{\tilde \phi}(k)}{\sigma^2k^2/2}\right],
\label{c1}
\end{equation}
where ${\tilde \phi}(k)$ is the Fourier transform of $\phi(\xi)$.
In particular, for the uniform distribution (example (iii)), one has ${\tilde 
\phi}(k)= \sin (k)/k$ and \eqref{cm1} gives
\begin{equation}
c = - \frac{1}{\pi}\, \int_0^{\infty} \frac{dk}{k^2}\, 
\ln \left[\frac{6}{k^2}\left(1-\frac{\sin
k}{k}\right)\right] = 0.29795219028\dots
\label{c2}
\end{equation}
The extraction of the constant correction term \eqref{c1}
explicitly from \eqref{psp1} 
turned out to be highly nontrivial and required a certain number
of delicate mathematical manipulations~\cite{CM}.
Interestingly, the same constant $c$ also appears in
an apparently different problem when one tries
to compute the average flux to a spherical trap in $3$-dimensions
of particles undergoing Rayleigh flights~\cite{Ziff}. The origin
of this connection has now been understood--both problems
are effectively described by exactly the same Wiener-Hopf integral equation,
albeit with two different initial conditions~\cite{MCZ}. Many other
interesting nontrivial exact results for this spherical trap problem  
have been recently computed in~\cite{MCZ,ZMC1,ZMC2}.

For the jump distributions where $\sigma^2$ is infinite, as in the case
of L\'evy flights, a similar formula for the expected maximum can be
derived~\cite{CM} from the Pollaczek-Spitzer formula. For example,
for ${\tilde \phi}(k)= 1- |ak|^\mu +O(k^2)$ (for small $k$ and with
$1<\mu\le 2$), the
expected maximum is given by~\cite{CM}
\begin{equation}
\frac{E(M_n)}{a}= \frac{\mu}{\pi} \,
\Gamma\left(1-\frac{1}{\mu}\right) n^{1/\mu} + \gamma
+O(n^{1/\mu-1})
\label{asymp2}  
\end{equation}
where the constant 
\begin{equation}
\gamma= \frac{1}{\pi}\int_0^{\infty} \frac{dz}{z^2} 
\ln {\left[\frac{1-{\tilde \phi}\left(\frac{z}{a}\right)}
{z^\mu}\right]}.
\label{clevy}
\end{equation}
For example, for ${\tilde \phi}(k)= \exp[-|ak|^\mu]$ with $1< \mu\le 2$, one
obtains~\cite{CM}
\begin{equation}
\gamma= \frac{1}{\pi}\int_0^{\infty} \frac{dk}{k^2} \ln
{\left[\frac{1-e^{-k^{\mu}}}{k^{\mu}}\right]}
=\frac{\zeta(1/\mu)}{(2\pi)^{1/\mu} \sin (\pi/{2\mu})}.
\label{special2}
\end{equation}
Note that for $0<\mu\le 1$, the expected maximum is strictly infinite.

We close this subsection by just pointing out another completely
different problem
where the expected maximum of a random walk plays an important
role. Recently we showed that the expected perimeter $\langle L_n\rangle$
of the convex hull of a $2$-dimensional random walk of $n$ steps
is exactly equal (up to a factor $2\pi$) to the expected maximum
of the $x$-components of this $2$-d random walk:
$\langle L_n\rangle = 2\pi \langle M_n\rangle$
where $M_n= {\rm max}(0,x_1,x_2,\ldots,x_n)$~\cite{hull1,hull2}.
This
connection allowed us to obtain a number of exact results
for the statistics of the convex hulls of random walks in two
dimensions. We do not discuss this problem in detail here, but
refer the interested readers to ~\cite{hull1,hull2} for details.

\subsection{Time at which the Random Walker's Trajectory Achieves its Maximum}

In the previous subsection we discussed the statistics of the maximum $M_n$
of an $n$-step walker. Another interesting question is the following:
given an $n$-step walker that started at the origin at step $0$, at which
step $m$ does the maximum $M_n$ happen? In other words, at which time step
the $n$-step walker is farthest (in the positive direction) from the 
origin. This time step $m$ of the occurrence of the maximum is itself
a random variable. It turns out that the probability distribution
of this time step $P(m|n)$ (given the total number of steps $n$ and
that $x_0=0$) is also closely related to the survival probability
$Q(0,n)$ discussed above. 

Before we discuss this, let us remark that for a continuous-time Brownian
motion \eqref{lange1} of total duration $t$ and starting at the origin,
the analogous probability density $P(t_m|t)$ of the time $t_m$ at which
the Brownian motion is maximally away from the origin in the positive 
direction was computed by L\'evy~\cite{Levy}
\begin{equation}
P(t_m|t) = \frac{1}{\pi\, \sqrt{t_m(t-t_m)}};\quad 0\le t_m\le t
\label{arcsin1}
\end{equation}
known as the celebrated L\'evy's arcsine law. The name `arcsine'
is due to the fact that
the 
cumulative 
distribution of $t_m$ has the arcsine form: ${\rm Prob}(t_m\le z 
t)=\frac{2}{\pi} \arcsin\left(\sqrt{z}\right)$ for $0\le z\le 1$.
Thus the maximum is more likely to occur at the begining $t_m=0$
or at the end $t_m=t$ of the time window, a fact slightly counterintuitive
given that the walk is symmetric around $0$.
Note that L\'evy's arcsine law also appears in the distribution
of the occupation time of a Brownian motion~\cite{Levy}. Let $t_+=\int_0^{t} 
\theta(x(\tau))\,d\tau$ be the time spent by a Brownian motion
of total duration $t$ on the positive side of the origin. Then
the probability density function of $t_+$ has
exactly the same form as in \eqref{arcsin1}
\begin{equation}
P(t_+|t)=\frac{1}{\pi\, \sqrt{t_+(t-t_+)}};\quad 0\le t_{+}\le t.
\label{arcsin2}
\end{equation}
This result looks rather simple, but again is nontrivial to derive.
For a derivation using Feynman-Kac path integral technique, see ~\cite{review}.

The two random variables $t_m$ and $t_+$ represent two rather different 
observables even though they share the same probability distribution.
The derivation in the two cases are also quite different. In mathematical
terms, one would say that $t_m\equiv t_+$ where $\equiv$ means that
these two random variables have the same statistical law. 
For the Brownian motion, one can prove this equivalence in law
directly~\cite{Feller}, 
without actually deriving the distribution separately in each case.
In fact, this equivalence between $t_m$ and $t_+$ holds 
for many other Markov processes as well~\cite{Feller}.

Coming back to the random variable $t_m$ of our interest, we note
that the distribution of $t_m$ has rather different shapes if one
puts various constraints on the Brownian motion.
For example, in case of a Brownian bridge i.e. a Brownian motion 
conditioned to
be at $x(0)=0$ and $x(t)=0$, the probability density of $t_m$ is
known to be uniform~\cite{Feller}
\begin{equation}
P(t_m|t)=\frac{1}{t};\quad\, 0\le t_m\le t.
\label{uniform1}
\end{equation}
Recently, using path integral methods, this distribution $P(t_m|t)$
was computed for a variety of other constrained Brownian motions, such
as Brownian excursion, Brownian meander, reflected Brownian bridge 
etc.~\cite{RFSM,MKRF,SMJP,Rajab}. Interestingly, $P(t_m=x|t=L)$
is also precisely the disorder-averaged equilibrium probability density
of a particle, moving in an external disordered potential in one dimension, at 
position $x$
in a box of size $L$~\cite{Rosso}.
Some of these results have been recently rederived by a 
functional renormalization group method~\cite{GSPL}. In addition, in the
context of the convex hull of Brownian motion in $2$-dimensions, it
turns out that to compute the mean area of the convex hull of
a $2$-d Brownian motion, one 
needs to compute the distribution $P(t_m|t)$ of the corresponding one 
dimensional Brownian motion~\cite{hull1,hull2}. Very recently,
the distribution $P(t_m|t)$ has been computed exactly~\cite{MRZ1} for
the random acceleration process (the continuous-time version
of the non-Markov evolution rule in \eqref{nonmarkov}). This, to my
knowledge, is perhaps the first exact result on $P(t_m|t)$ for
a non-Markov process. 

\begin{figure}
\includegraphics[height=10.0cm,width=14.0cm,angle=0]{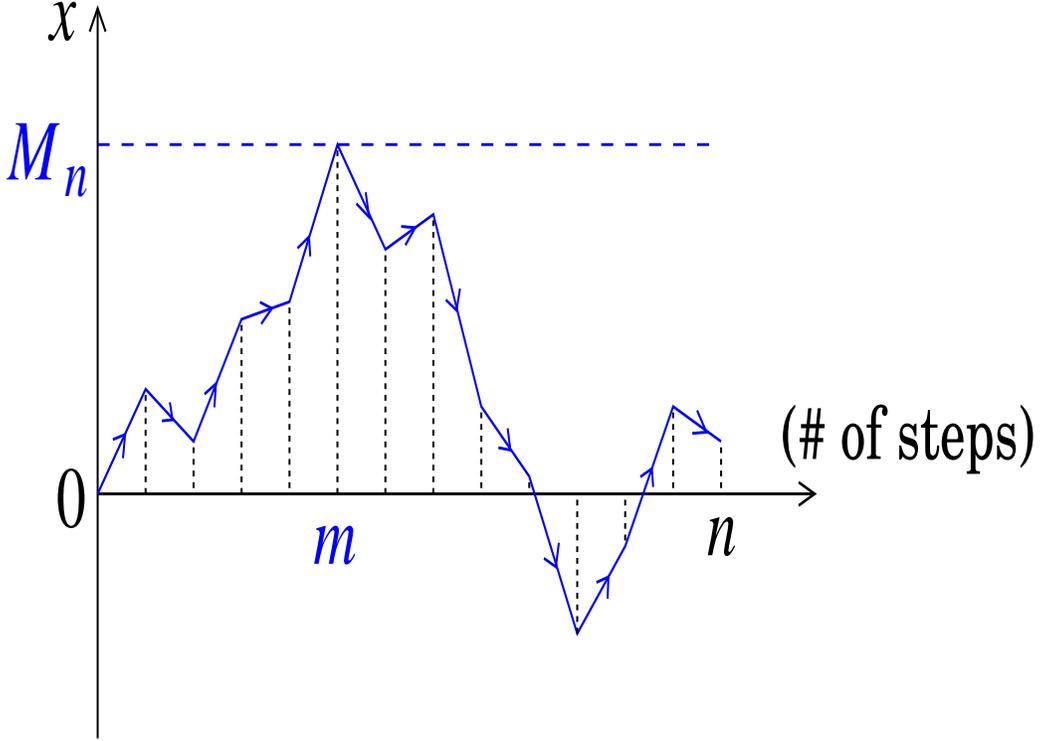}
\caption{\label{fig:rw2} A trajectory of a random walker of $n$ steps,
starting at the initial position $0$, achieving its maximum $M_n$ at an
intermediate step 
$m$.}
\end{figure}

The analogous distribution $P(m|n)$ for the discrete-time random walk
process in \eqref{evol1} for arbitrary continuous and symmetric jump length 
distribution
$\phi(\xi)$ can be computed exactly from the knowledge of the survival
probability $q(n)=Q(0,n)$. To see this, consider Fig. \ref{fig:rw2}
Let us just invert this figure and look at the trajectory from the
position $M_n$, i.e., make a change of variable: $z_k=M_n-x_k$.
Next we decompose the trajectory into two parts: the left side for
time steps between $0$ and $m$ and the right side for time steps between $m$ 
and $n$. Using the Markov property, these two parts are independent
of each other. In the inverted picture, for the left side,
let us also invert the `time', i.e., propagate backwards. 
One has to thus consider 
all $z_k$ paths that start at $z=0$ and stays positive up to $m$ steps
(which is equivalent to saying that $x_k$'s stays below $M_n$). 
Note that finally we have to integrate over all possible $M_n$
which means in the inverted picture the final value of $z_m$
is integrated over.
Thus the 
contribution from this left part is just $q(m)=Q(0,m)$. A similar
reasoning shows that the
contribution from the right part is $q(n-m)=Q(0,n-m)$.
Multiplying one gets, upon using the Sparre Andersen result \eqref{SA2},
\begin{equation}
P(m|n)= q(m)q(n-m)= {{2m} \choose {m}}{{2(n-m)} \choose 
{(n-m)}}\, 2^{-2\,n}.
\label{dist1}
\end{equation}
One can check easily that this distribution is normalized to unity: 
$\sum_{m=0}^n P(m|n)=1$. Amazingly, thanks to the Sparre Andersen
result, the distribution $P(m|n)$ is again universal for all $m$ and $n$, 
i.e., independent of the jump length distribution as long as it is continuous.
Thus, it is given by the same formula \eqref{dist1} for Gaussian, uniform
or even for L\'evy flights!

In the limit of large $m$ and $n$ (keeping the ratio $m/n=x$ fixed), one gets
\begin{equation}
P(m|n) \simeq \frac{1}{\pi \sqrt{m(n-m)}}
\label{dist2}
\end{equation}
which, once again, may naively look like the arcsine law for the Brownian 
motion \eqref{arcsin1}. However, note that this asymptotic result
in \eqref{dist2} is valid even for L\'evy flights. This `arcsine'
looking law, valid for arbitrary distribution, is not quite the
same as the `arcsine' law in the Brownian limit for the same reason
discussed before in the context of the survival probability.

\section{Second Application: Statistics of Records}

In this section we will discuss
another beautiful recent application of the Sparre Andersen theorem \eqref{SA2}
that results in the universal statistics of records in a random walk
sequence (including the L\'evy flights)~\cite{MZ}.
Statistics of records forms an integral
part of diverse fields including
meteorology~\cite{climate,RP}, hydrology~\cite{Matalas}, 
economics~\cite{Barlevy},
sports~\cite{GTS,Glick,BRV} and entertainment
industries among
others.
In popular media such as television or newspapers, one always hears and reads about 
record
breaking events. It is no
wonder that {\em Guinness Book of Records} has been a world's best-seller since 1955.
Understanding the statistics of records is particularly important in   
the
context of current issues of climatology such as global warming.

Consider any discrete time series  $\{x_0,x_1,x_2,\ldots,x_n\}$ of $n$ entries 
that 
may
represent, e.g.,
the daily temperatures in a city or the stock prices of a company or the budgets
of Hollywood films. 
A {\em record} happens at step $i$ if the $i$-th entry $x_i$
is bigger than all previous entries $x_0$, $x_1$, $\ldots$, $x_{i-1}$.
Statistical questions that naturally arise are: (a) how many records occur up 
to
step
$n$? (b) How long does a record survive? (c) what is the age of 
the longest surviving record? Answering these questions 
is the main goal of the theory of records.

The mathematical theory of records has been studied for over 50
years~\cite{Chandler,Nevzorov,ABN,SZ} and the questions posed in the previous
paragraph are well understood in
the case when $x_i$'s are 
i.i.d random variables.
Recently, there has been a resurgence of interest
in the record theory due to its multiple applications in diverse
complex systems such as spin glasses~\cite{SG}, adaptive processes~\cite{Orr} and
evolutionary models of biological
populations~\cite{Krug1,Evol} and models of growing networks~\cite{GL}. The results 
in the record theory of i.i.d variables have
been rather useful in these different contexts.
Recently, Krug has studied the record statistics when the entries have non-identical  
distributions but still retaining their independence~\cite{Krug2}.
However, in most realistic situations the entries of the time series are
{\em correlated}.
Very little seems to be known about the statistics of records for a correlated 
time series. 
Recently, we developed a general formalism~\cite{MZ} to
study the statistics of 
records in a random walk sequence
evolving via \eqref{evol1} with an arbitrary jump distribution
$\phi(\xi)$. We showed~\cite{MZ} that for symmetric and continuous jump 
distributions,  
the statistics of records 
have universal properties as a
consequence of the Sparre Andersen theorem discussed before.
Below we discuss this formalism developed in \cite{MZ} in
some details.

\begin{figure}
\includegraphics[width=.9\hsize]{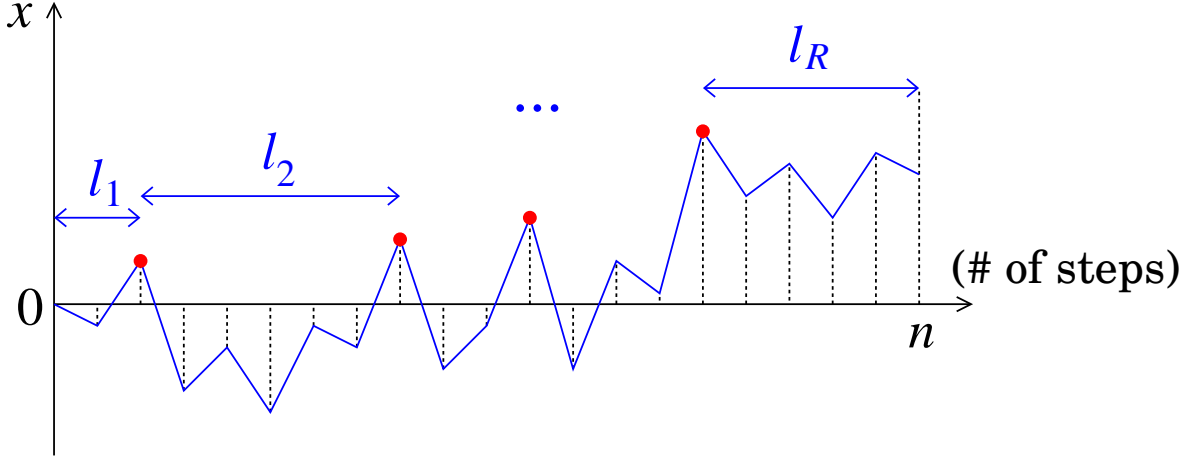}
\caption{A realization of the random walk sequence 
$\{x_0=0,x_1,x_2,\ldots, x_n\}$ of
$n$
steps with $R$ records.
Records are shown as big red dots. Note that a local maximum
of the walk is not necessarily a record. The set $\{l_1,l_2,\ldots, 
l_R\}$ denotes the time 
intervals
between successive records.}
\label{fig:record1}
\end{figure}

To proceed, let us consider a realization of the sequence $x_i$'s  
in \eqref{evol1} up to $n$ steps. The discussion below
is general and holds even for asymmetric jump distribution $\phi(\xi)$.
Let $R$
be the
number of records in this realization. 
We use the convention that the first entry $x_0$ is counted
as a record.
Evidently $R$ is an integer. Let $l_i$ denote the time interval
between the $i$-th and the $(i+1)$-th record. Thus, $l_i$
is the age of the $i$-th record, i.e., it denotes the time up to
which the $i$-th record survives. We will
use the shorthand notation ${\vec l}=\{l_1,l_2,\ldots, 
l_R\}$ to denote the set of $R$ successive intervals (see Fig. 
\ref{fig:record1}).
Note that the last record, i.e., the $R$-th record
still stays a record at the $n$-th step since there is no more record breaking   
events after it. Hence $l_R$ (the last one in Fig. \ref{fig:record1})
denotes the number of steps after the occurrence of the last record
till the last step $n$. 
The main idea is to first calculate the joint probability distribution
$P\left(\vec l, R|n\right)$ of the ages $\vec l$ and 
the number $R$ of records, given
the
length $n$ of the sequence.  

To compute this joint distribution
we need two quantities as inputs.
First, let $q_{-}(l)$ denote the probability that a walk, starting initially at 
$x_0$,
stays {\em below} its starting position $x_0$ up to step $l$.
Clearly $q_{-}(l)$ does not depend on the starting position $x_0$ due
to translational invariance 
and one can just set $x_0=0$. Then $q_{-}(l)$ is precisely
the survival probability defined in \eqref{downsurv}
whose generating function ${\tilde q}_{-}(s)$ is given by the generalized
Sparre Andersen result in \eqref{gsadown}.
Recall that for the symmetric case $q_{-}(l)=q_{+}(l)=q(l)$ is 
universal and its generating
function is given exactly in \eqref{SA1}
\begin{equation}
{\tilde q}(s)=\sum_{l=0}^\infty q(l)\, s^l = \frac{1}{\sqrt{1-s}}.
\label{qgf1}
\end{equation} 

The second input is the first-passage probability $f_{-}(l)$ that the walker 
crosses 
its
starting
point $x_0$ for the first time between steps $(l-1)$ and $l$ from
below $x_0$ (see Fig. \ref{fig:record1}). Once again, 
$f_{-}(l)$ 
does not 
depend on the starting point $x_0$ due to translational invariance and 
and one can set $x_0=0$. Setting $x_0=0$, it follows
that $f_{-}(l)= q_{-}(l-1)-q_{-}(l)$ whose generating function
can be expressed in terms of that of $q_{-}(l)$ 
\begin{equation}
{\tilde f}_{-}(s)= \sum_{l=1}^{\infty} f_{-}(l) s^l = 1-(1-s){\tilde 
q}_{-}(s).
\label{fgf0}
\end{equation}
In the symmetric case, $f_{+}(l)=f_{-}(l)=f(l)$ with
a generating function
\begin{equation}
{\tilde f}(s) =1-(1-s){\tilde
q}(s)= 1-\sqrt{1-s}
\label{fgf1}
\end{equation}
where we have used \eqref{qgf1}.
 
Armed with these two ingredients $q_{-}(l)$ and $f_{-}(l)$, we can then write 
down
explicitly the joint distribution of the ages $\vec l$ and the number $R$ of records
\begin{equation}
P\left(\vec l, R|n\right)= f_{-}(l_1)\,f_{-}(l_2)\,\ldots 
f_{-}(l_{R-1})\,q_{-}(l_R)\,
\delta_{ {\sum_{i=1}^R l_i,\, n}}
\label{joint1}
\end{equation}
where we have used the Markov renewal property of random walks which dictates that
the successive intervals are statistically independent, except for the
global sum rule that the total interval length is $n$ (see Fig. 
\ref{fig:record1}) which
is incorporated by the delta function. Note that since the $R$-th record
is the last one (i.e., no more records have happened after it),
the interval to its right has distribution $q_{-}(l)$ rather
than $f_{-}(l)$. One can check that $P\left(\vec l, R|n\right)$ is normalized 
to unity
when summed over $\vec l$ and $R$. 

Note that in the case of symmetric jump distribution,
since $q_{-}(l)=q(l)$ and $f_{-}(l)=f(l)$ are universal   
due to the symmetric Sparre Andersen theorem, it follows that $P\left(\vec l, 
R|n\right)$
and all marginals of it are also universal. Below we will focus
on the symmetric case only.

\subsection{ Universal Distribution of the Number of Records up to step $n$}

Let us focus here on the case of symmetric jump distribution 
$\phi(\xi)=\phi(-\xi)$ where $\phi(\xi)$ is continuous.
In this case we can replace $q_{-}(l)$ by $q(l)$ and $f_{-}(l)$
by $f(l)$ in the joint distribution \eqref{joint1}.
Let us first compute the probability distribution of the number of 
records $R$, $P(R|n)=\sum_{\vec
l} P\left(\vec l, R|n\right)$. To perform this sum, it is easier to consider its
generating
function. Multiplying \eqref{joint1} by $s^n$ and summing over $\vec l$, one
gets
\begin{equation}
\sum_{n=R-1}^{\infty} P(R|n) s^n= [{\tilde f}(s)]^{R-1}
{\tilde q}(s)=\frac{(1-\sqrt{1-s})^{R-1}}{\sqrt{1-s}}
\label{nore2}
\end{equation}
where we have used the explicit expressions for ${\tilde q}(s)$ and ${\tilde 
f}(s)$ from Eqs. \eqref{qgf1} and \eqref{fgf1}.

By expanding in powers of $s$ and computing the coefficient of $s^N$ one gets
the explicit result~\cite{MZ}
\begin{equation}
P(R|n)= {{2n-R+1}\choose n}\, 2^{-2n+R-1}
\label{nore1}
\end{equation}
which is {\em universal} for all $R$ and $n$.
The moments of $R$ are also naturally universal
and can be computed for all $n$. 
For example, the first three moments are
\begin{eqnarray}
\langle R\rangle &= &(2n+1)\,{2n \choose n}\, 2^{-2n} \nonumber \\ 
\langle R^2\rangle &=& 2n+ 2 - \langle R\rangle \nonumber \\
\langle R^3\rangle &=& -6 n - 6 + (7 + 4n) \langle R\rangle.
\label{mom2}    
\end{eqnarray}
In particular, for large $n$, the mean,
variance and the skewness behave as
\begin{eqnarray}
{\rm Mean} &:& \quad\quad \langle R\rangle \simeq   
\frac{2}{\sqrt{\pi}}\sqrt{n} 
\nonumber \\
{\rm Variance} &:&  \quad\,\, \langle R^2\rangle-{\langle R\rangle}^2 \simeq  
2\left(1-\frac{2}{\pi}\right)\, n \nonumber \\
{\rm Skewness} &:& \quad\,\, \frac{\langle \left(R-\langle 
R\rangle\right)^3\rangle}{ {\langle \left(R-\langle 
R\rangle\right)^2\rangle}^{3/2} }
\simeq  \frac{4(4-\pi)}{(2\pi-4)^{3/2}}  
\label{mom1}
\end{eqnarray}
In ~\cite{MZ}, these results were also verified numerically 
for different jump length distributions (uniform, Gaussian, Cauchy)
all giving the same universal answer.

The results in \eqref{mom1} suggest that there is only
a single scale for the number of records $R\sim n^{1/2}$.
This is confirmed by analysing the full distribution $P(R|n)$ of $R$
in \eqref{nore1} in the limit of large $n$. One finds that $P(R|n)$
actually has the following scaling form for large $n$~\cite{MZ}
\begin{equation}
P(R|n) \simeq \frac{1}{n^{1/2}}\,g\left(\frac{R}{n^{1/2}}\right);\quad\quad 
g(x)= \frac{1}{\sqrt{\pi}}\, e^{-x^2/4}.
\label{rdist}
\end{equation}
Thus the distribution is broad in the sense that the mean and the standard 
deviation measuring the fluctuation around the mean, both scale as $\sim 
n^{1/2}$. Also, the mode of this distribution, i.e., the
most probable (typical) value of $R$ is at $R=0$. It is interesting to 
compare this result for the random walk 
sequence \eqref{evol1} with  
that of an uncorrelated i.i.d sequence where each entry $x_i$ is a random 
variable
drawn from some distribution $p(x)$. In the latter case, it is well
known~\cite{Nevzorov} that the distribution of the number of records $P(R|n)$ does 
not
depend on $p(x)$, and for large $n$, it approaches a Gaussian,
\begin{equation}
P(R|n)\simeq \frac{1}{\sqrt{2\pi \log n}}\,\exp\left[-\frac{(R-\log n)^2}{2\log 
n}\right]
\label{rdistgauss}
\end{equation}
with mean $\langle R\rangle = \log n$ and
the standard deviation $\sqrt{\log n}$. 
This distribution has its peak at $R=\log n$, in
stark contrast to the random walk case where the
most probable value of $R$ is zero. In addition,
even the fluctuations
of $R$ are {\em small} compared to the mean for large $n$,
again in contrast to the random walk case where the
fluctuations are large $\sim O(\sqrt{n})$ for large $n$.
Thus the effect of correlation in the random walk sequence
manifests itself in a broad scaling distribution for the
number of records.

\subsection{ Universal Age Distribution of Records}

Since the mean number of records grows as $\langle R\rangle \sim n^{1/2}$, it follows 
that the {\em typical} age of a record grows also as $\langle l\rangle\sim n/\langle
R\rangle\sim n^{1/2}$ for large $n$.
However there are {\em rare} records that are not typical and their
ages follow different statistics. For example, what is age distribution
of the longest lasting and the shortest lasting records? These {\em extreme}
statistics of ages can also be derived from the joint distribution
in \eqref{joint1} and hence they are also universal and independent of 
$\phi(\xi)$.

Let us first consider the longest lasting record with age $l_{\rm max}= {\rm
max}(l_1,l_2,\ldots, l_R)$. It is easier to compute its cumulative distribution
$Y(l|n)={\rm Prob}[l_{\rm max}\le l]$ given $n$.
Now, if $l_{\rm max}\le l$, it follows that each of the intervals $l_i\le l$ 
for
$i=1,2,\ldots, R$. Thus, we need to sum up \eqref{joint1} over all $l_i$'s
and $R$ such that $l_i\le l$ for each $i$. As usual it is easier 
to carry out this    
summation
by considering the generating function and we get
\begin{equation}
\sum_n Y(l|n)\, s^n = \frac{\sum_{l'=1}^l q(l') s^{l'}}{1- \sum_{l'=1}^l f(l') 
s^{l'}}.
\label{fngf1}
\end{equation}
One can extract, in principle,
the distribution $Y(l|n)$ from this general expression.
In particular, the asymptotic large $n$ behavior of the 
average $\langle l_{\rm max}\rangle= \sum_{l=1}^{\infty} [1-Y(l|n)]$ 
can be extracted explicitly~\cite{MZ}
\begin{equation}
\langle l_{\rm max}\rangle\simeq c_1\, n; \quad\quad\quad c_1=2\int_0^{\infty} 
dy\, 
\log 
\left[1+ 
\frac{1}{2\sqrt{\pi}}
\,\Gamma(-1/2,y)\right]= 0.626508\ldots
\label{cons1}
\end{equation}
where $\Gamma(-1/2,y)= \int_y^{\infty}dx\, x^{-3/2}\, e^{-x}$
is the incomplete Gamma function.
Thus, the age of the longest record ($\sim n$)
is much large than the typical age ($\sim \sqrt{n}$) for large $n$.

For the shortest lasting record $l_{\rm min}= {\rm min}(l_1,l_2,...l_R)$, it is
also useful to consider the cumulative distribution $Z(l|n)={\rm Prob}[l_{\rm 
min}\ge l]$ given $n$.
This event is equivalent
to having the lengths, $l_i\ge l$ for all $i=1,2,\ldots R$.
Following similar procedure as in the case of the longest lasting record,
one finds the generating function
\begin{equation}
\sum_n Z(l|n)\, s^n = \frac{\sum_{l'=l}^{\infty} q(l') s^{l'}}{1- 
\sum_{l'=l}^\infty 
f(l) s^{l'}}.
\label{gngf1}
\end{equation}
One can then extract, in a similar way, the asymptotic large $n$ behavior
of $\langle
l_{\rm min}\rangle\sim \sqrt{n/\pi}$~\cite{MZ}. Thus, the mean age of the 
shortest lasting
record grows in a similar way as that of a typical record, i.e., as $\sqrt{n}$,
albeit with a smaller prefactor $1/\sqrt{\pi}=0.56419\ldots$ compared with
$\sqrt{\pi/4}=0.88623\ldots$.

\subsection{Two Generalizations}

In the discussion above for the statistics of records, we had assumed 
that the jump length distribution
$\phi(\xi)$ is symmetric and continuous. However, the basic renewal
equation \eqref{joint1} is valid for continuous but asymmetric
jump distribution as well. The only difference is
that we have to use the appropriate expressions
for $f_{-}(l)$ and $q_{-}(l)$ from the generalized Sparre
Andersen theorem. For example, the generating function
for the distribution $P(R|n)$ for the number
of records up to step $n$ is given by
the asymmetric version of \eqref{nore2}
\begin{equation}
\sum_{n=R-1}^{\infty} P(R|n) s^n= [{\tilde f}_{-}(s)]^{R-1}\,
{\tilde q}_{-}(s)= \left[1-(1-s) {\tilde q}_{-}(s)\right]^{R-1}\,{\tilde 
q}_{-}(s)
\label{nore3}
\end{equation} 
where ${\tilde q}_{-}(s)$ is given by \eqref{gsadown}.

Indeed, for the special case of a random walk sequence
in presence of a drift $\mu$ and Cauchy distributed jumps
as in \eqref{drift}, one can obtain
explicit results~\cite{LW} for $P(R|n)$ in \eqref{nore3}. 
Using
\eqref{cauchy4} one gets the exact generating function: ${\tilde q}_{-}(s)= 
(1-s)^{-\zeta_{-}}$ and substituting this in \eqref{nore3}
gives 
\begin{equation}
\sum_{n} P(R|n)\, s^n = 
\frac{\left[1-(1-s)^{1-\zeta_{-}}\right]^{R-1}}{(1-s)^{\zeta_{-}}}
\label{cauchy6}
\end{equation}
from which it follows that the average number of records grows anomalously
$\langle R\rangle \sim n^{1-\zeta_{-}}\sim n^{\zeta_{+}}$ for large $n$. 
Using $\zeta_{+}= 1/2+ {\tan}^{-1}(\mu/a)/\pi$, one sees that
as $\mu\to \infty$ (positive drift away from the origin), $\zeta_{+}\to 1$
and thus the
average number of records
grows linearly with the number of steps $n$, i.e., at every step
a new record happens on an average. Of course, this is expected in presence
of an infinite drift since the particle moves ballistically 
in the positive semi-axis. On the other hand, as $\mu\to -\infty$,
$\zeta_{+}\to 0$ indicating that the average number of records
do not grow with $n$ for large $n$. This is also expected since
the particle mostly stays on the negative side of the origin when
$\mu\to -\infty$ and thus hardly ever makes a positive record.
These 
results were then used
to understand the anomalous avalanche size distribution in a model of a 
particle
moving in a random potential~\cite{LW}.
 
\begin{figure}
\includegraphics[height=10.0cm,width=14.0cm,angle=0]{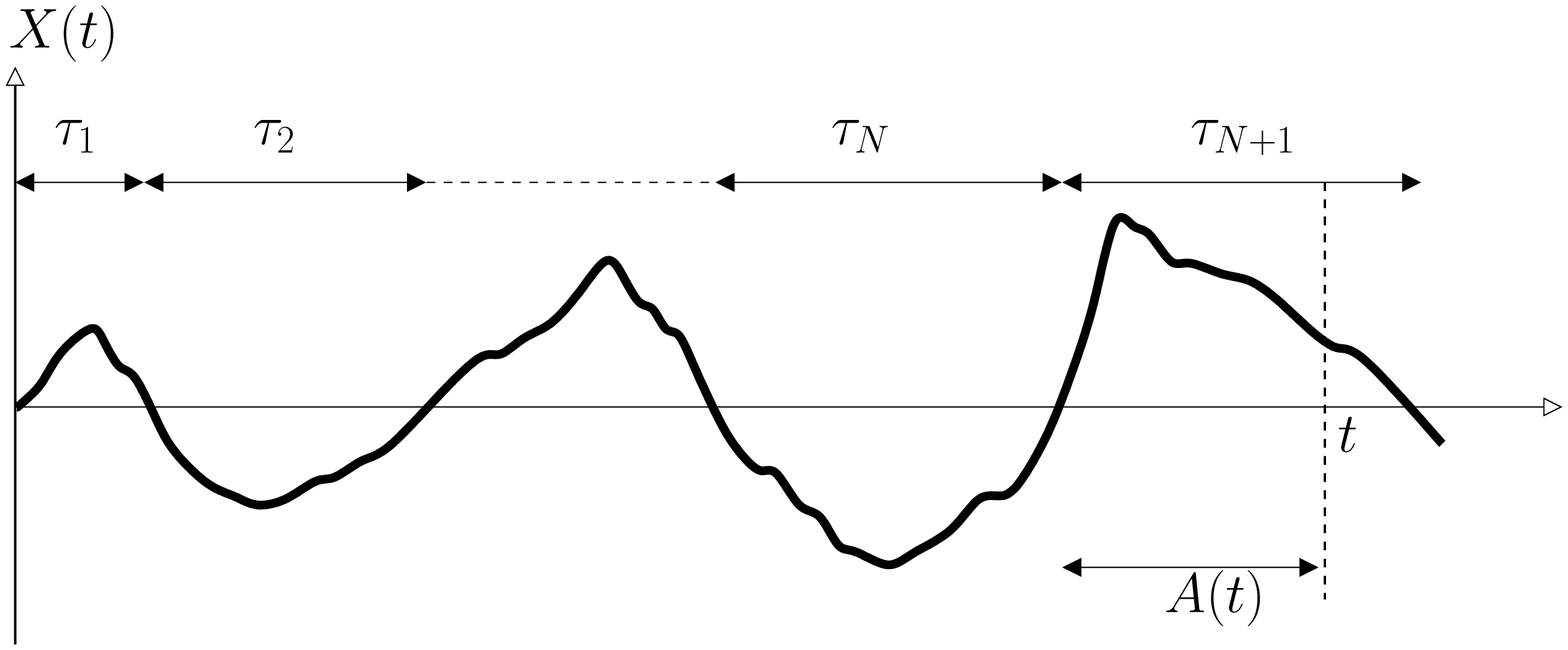}
\caption{\label{fig:cartoon} A trajectory of a Brownian motion
over [0,t] with $N$ completed excursions of lengths $[\tau_1, 
\tau_2,\ldots,\tau_N]$ and the last incomplete excursion of
length $A(t)$.}
\end{figure}

Another interesting generalization of these results emerged from
the following observation: it turns out that the constant $c_1=
0.626508\ldots$ that appears as the prefactor of the
linear growth of the longest lasting record in \eqref{cons1}
also appears in the excursion theory of Brownian motion~\cite{PY}.
Let us consider a Brownian motion over a time interval $[0,t]$
and consider the set of successive zero crossing intervals or excursions (see
Fig. \ref{fig:cartoon}).
Let us denote the maximum excursion length up to time $t$ by $l_{\rm max}(t)$
\begin{equation}
\label{def_lmax}
l_{{\rm max}}(t) = {\max} ( \tau_1, \tau_2, \cdots, \tau_N, A(t)) \;
\end{equation}
where $A(t)$ denotes the length of the last interval before $t$ (see
Fig. \ref{fig:cartoon}).
Let $Q(t)= {\rm Prob}[l_{\rm max}(t)=A(t)]$ denote the probability
the last incomplete excursion is the longest one.
Then it turns out~\cite{PY} that $Q(t)$ tends, for large $t$, to the same 
constant
$Q(t)\to c_1=0.626508\ldots$ as in Eq. \eqref{cons1}.
We were able to 
understand recently
why this same constant $c_1$ appears in apparently different observables
namely (i) in the length of the longest lasting record and (ii) the 
probability $Q(t)$ that the last excursion is the longest~\cite{GMS}.
This understanding led us to study the statistics of $l_{\rm max}(t)$
and that of $Q(t)$ for generic stochastic processes going beyond
the simple Brownian motion~\cite{GMS}.
The statistics
of $l_{\rm max}(t)$ turns out to have interesting
universal features that allowed us to distinguish between stochastic processes that
are smooth (i.e. with a finite density of zero crossings) 
versus the ones that are rough (where the density of zero crossings
is infinite as in the case of the Brownian motion)~\cite{GMS}.

\section{Summary and Conclusion}

In summary, I have discussed the universal first-passage properties associated
with a discrete-time random walk sequence consisting of $n$ steps, where
the walker starts at the origin $x_0=0$ and at each step jumps by
a random amount drawn independently at each step from a symmetric
and continuous distribution $\phi(\xi)$. The first-passage probability
is universal, i.e., independent of the jump length distribution
due to the Sparre Andersen theorem. We have then used the consequence
of this result on the statistics of two extreme random variables: (i)
the global maximum of the walk and the step at which it occurs and (ii)
the number and ages of records. We have seen that the distribution
of the time of the maximum as well as the record statistics become
universal as a consequence of the Sparre Andersen theorem. 

The distribution
of the value of the maximum, however, is non-universal and depends
explicitly on $\phi(\xi)$. 
The random variables belonging to this sequence are correlated.
For the distribution of the maximum, the standard EVS of i.i.d.
random variables does not apply due to these correlations. 
The computation of the distribution of the maximum for this
discrete-time sequence is thus nontrivial due to these correlations, even 
though in the corresponding
continuous-time Brownian motion it is easy to compute. However, thanks
to the Pollaczek-Spitzer formula, one knows, at least in principle, how
to compute the generating function of this maximum distribution
for arbitrary symmetric and continuous $\phi(\xi)$. 
The leading large $n$ behavior of the moments of the maximum can
be extracted relatively easily from this explicit Pollaczek-Spitzer
formula. However, extracting the subleading finite size correction term
turns out to be much trickier. 
At least for the
expected maximum, we have seen how to compute exactly the leading 
finite size correction
term, for the case when the jump distribution has
a finite variance and also for the case of L\'evy flights with index $1<\mu\le 
2$.
These results are interesting because the expected maximum of a discrete-time
random walk is exactly related to the perimeter of the convex hull of 
a planar random walk which has important applications in the estimation
of home range of animals in ecology~\cite{hull1,hull2}.

It would also be interesting to
compute the expected maximum and the distribution of the time of its
occurrence in presence of a drift. In the Brownian limit, in presence
of a drift, the distribution
of the time at which the maximum occurs has been computed using a
path integral method~\cite{SMJP}, with interesting applications in finance. 
However, for the discrete-time case,
I am not aware of any result so far and it would be interesting to compute this
distribution.

There are interesting generalizations of the results presented here.
For example, concerning the statistics of records, we have studied
only the statistics of `positive' records, i.e., when
the value $x_i$ of a record that occurs at step $i$ is bigger
than all previous values, given that the sequence started at $x_0=0$.
It would be interesting to investigate the statistics of the records of 
the absolute values
of the sequence, i.e., of $\{0,|x_1|,|x_2|,\ldots, |x_n|\}$ which, to my
knowledge, has not yet been studied~\cite{Krugquestion}.

As I already mentioned, the record statistics of this Markov sequence has been
studied in presence of a constant drift
with interesting
applications in avalanche dynamics~\cite{LW}. In particular,
we have seen one case, namely
the Cauchy distribution with drift, where the average number of records
grows with the sequence size $n$ anomalously with a nontrivial
drift-dependent exponent~\cite{LW}. 
It would not be difficult to compute the distribution of the ages
of records in this particular case. The study of the age distribution
of records for arbitrary asymmetric jump distribution remains
an open problem.

Another interesting generalization is to consider the Markov
sequence generated by the recursion: $x_n=r\, x_{n-1}+\xi_n$
where $0\le r \le 1$ is a parameter and $\xi_n$'s are, as before,
symmetric i.i.d. noise variables. This is just a discrete-time
analogue of the continous-time Orstein-Uhlenbeck (OU) process
of a particle moving in a harmonic potential. This is seen
by writing, $x_n-x_{n-1}=-(1-r)x_{n-1}+\xi_n$ which, in the
continuous-time limit (alongwith $r\to 1$ limit), becomes the process OU process, $dx/dt = 
-\lambda 
x+\xi(t)$ where $\xi(t)$ is a zero mean Gaussian white noise.
This discrete-time sequence has many applications, e.g., it appears
in the context of the practical sampling of experimental data
on the persistence of a stochastic process~\cite{MBE1,MBE2,surface,practice}
and also in the simple system of a ball bouncing non-elastically
on a noisy platform~\cite{MK}. In the latter context, the parameter
$0\le r\le 1$ represents the coefficient of restitution of
the collision of the ball with the platform~\cite{MK} and
the Brownian limit $r=1$ corresponds to elastic collision.
The first-passage properties of this sequence for generic $0<r<1$ turns out to
be highly nontrivial even for a Gaussian noise distribution~\cite{MBE1}. 
Explicit exact result is known only for the
exponential noise distribution~\cite{MK}. While, for generic $0<r<1$, these
first-passage properties are nonuniversal, one recovers
several interesting universal properties in the elastic limit $r\to 
1$~\cite{MK}. It would be interesting to study the statistics of the
maximum and that of the records in this simple Markov sequence for arbitrary 
$0< r<1$ and arbitrary
noise distribution.

In conclusion, there are still many unresolved questions associated
with even simple one dimensional random walks. Depending on the new 
applications, new questions 
emerge requiring new techniques to solve them which are often nontrivial
and interesting.

\vspace{0.4cm}

\noindent{\bf Acknowledgements:} The results presented here are based on my 
joint 
work with A. Comtet and R.M. Ziff. It is a great pleasure to thank them.
I acknowledge useful discussions on related topics with J.-P. Bouchaud, A.J. 
Bray,  
D. Dhar, C. Godr\'eche, P. Le Doussal, M.J. Kearney, J. Randon-Furling, A. 
Rosso, 
G. Schehr, C. Texier and
M. Yor.
I also thank J. Krug and S. Redner for many useful 
discussions on the statistics of records.  
Some of the general topics covered in my lectures here have some
overlap with the lectures given by J.-P. Bouchaud, P. Le Doussal
and myself at the
Beg-Rohu summer school (2008). The lecture notes
of this Beg-Rohu school, compiled from the notes of M. Lenz, are available
at the web site: http://ipht.cea.fr/Meetings/BegRohu2008/notes.html. 
I take this opportunity to thank M. Lenz for his nice notes.

\end{document}